\documentclass[prd,twocolumn,a4paper,superscriptaddress,floatfix]{revtex4}
\usepackage{graphicx}
\usepackage{amssymb, amsmath}
\begin{document}


\title{Constraining the helium abundance with CMB}

\author{Roberto Trotta}
\email[Electronic address: ]{roberto.trotta@physics.unige.ch}
\affiliation{D\'epartement de Physique Th\'eorique, Universit\'e de
Gen\`eve, 24 quai Ernest Ansermet, CH-1211 Gen\`eve 4,
Switzerland}

\author{Steen H. Hansen}
\email[Electronic address: ]{hansen@physik.unizh.ch}
\affiliation{Physik Institut, Winterthurerstrasse 190,
8057 Zurich, Switzerland}

\begin{abstract}
  We consider for the first time the ability of present-day cosmic
  microwave background (CMB) anisotropies data to determine the
  primordial helium
  mass fraction, $Y_p$. We find that CMB data alone gives the
  confidence interval $0.160 < Y_p < 0.501$ (at 68\% c.l.).  We analyse the
  impact on the baryon abundance as measured by CMB and discuss the
  implications for big bang nucleosynthesis.  We identify and discuss
  correlations between the helium mass fraction and both the redshift
  of reionization and the spectral index. We forecast the
  precision of future CMB observations, and find that Planck alone will
  measure $Y_p$ with error-bars of $5\%$. We point out that the
  uncertainty in the determination of the helium fraction will have
  to be taken into account in order to correctly estimate the baryon
  density from Planck-quality CMB data.
\end{abstract}
\maketitle

\def\gsim{\;\raise0.3ex\hbox{$>$\kern-0.75em\raise-1.1ex\hbox{$\sim$}}\;}
\def\lsim{\;\raise0.3ex\hbox{$<$\kern-0.75em\raise-1.1ex\hbox{$\sim$}}\;}
\newcommand{\ie}{{\it i.e.~}}
\newcommand{\ob}{\omega_b}
\newcommand{\OLa}{\Omega_\Lambda}
\newcommand{\zreion}{z_r}
\newcommand{\separator}[1]{\multicolumn{6}{|c|}{#1} \\\hline}
\newcommand{\kA}{\mathcal{A}}
\newcommand{\kB}{\mathcal{B}}
\newcommand{\kV}{\mathcal{V}}
\newcommand{\kR}{\mathcal{R}}
\newcommand{\kM}{\mathcal{M}}
\newcommand{\rZ}{\mathcal{T}}
\newcommand{\fsky}{f_{\textrm{sky}}}
\newcommand{\xres}{x_e^{\textrm{res}}}
\newcommand{\fres}{f_e^{\textrm{res}}}
\newcommand{\tres}{\tau^{\textrm{res}}}
\newcommand{\zdec}{z_{\textrm{dec}}}
\newcommand{\treion}{\tau_{\textrm{r}}}

\section{Introduction}
\label{introduction}
Our understanding of the baryon abundance has increased dramatically
over the last few years. This improvement comes from two independent
paths, namely big bang nucleosynthesis (BBN) and cosmic microwave
background radiation (CMB). Absorption features from high-redshift
quasars allow to measure precisely the deuterium
abundance, D/H. Combined with BBN calculations, this provides a reliable
estimate of the baryon to photon ratio, $\eta$.
An independent determination of the baryon content of
the universe from CMB anisotropies comes from the increasingly precise
measurements of the acoustic peaks, which bear a characteristic
signature of the photon-baryon fluid oscillations. The agreement
between these two completely different approaches is both remarkable
and impressive (see details below).  The time is therefore ripe to
proceed and test the agreement between other light elements which are
also probed both with BBN and CMB.

The helium abundance has been measured for many years from
astrophysical systems. However, the error-bars are seemingly dominated
by systematic errors which are hard to assess.  Fortunately, the
dependence of the helium mass fraction on the CMB anisotropies
provides an independent way to measure $Y_p$.  The aim of this work is
to present the first determination of the helium abundance from CMB
alone, and to clarify the future potential of this method.  The latest
CMB data are precise enough to allow taking this further step, and in
view of the emerging ``baryon tension'' between BBN predictions from
observations of different light elements~\cite{olive03} possibly requires taking such
a step. The advantage of using CMB anisotropies rather than the
traditional astrophysical measurements, is that CMB provide a clear
measurement of the primordial helium fraction before it could be
changed by any astrophysical process. On the other hand the dependence
of the CMB power spectrum on $Y_p$ is rather mild, a fact which makes
it presently safe to fix the value of the helium mass fraction with
zero uncertainity for
the purpose of CMB data analysis of other cosmological parameters.

The paper is organized as follows. In section~\ref{sec:bbn} we review
the standard Big Bang Nucleosynthesis scenario. Section~\ref{sec:cmb}
discusses the role of the helium mass fraction for cosmic microwave
background anisotropies, the methods used and results. We discuss our
forecast for future CMB observations in section~\ref{sec:future}, and offer our conclusions in
section~\ref{sec:conclusions}

\section{Big Bang Nucleosynthesis}
\label{sec:bbn}

\subsection{The standard scenario}

The standard model of big bang nucleo\-syn\-the\-sis (SBBN) has only one
free parameter, namely the baryon to photon ratio $\eta_{10} =
n_b/n_\gamma10^{10}$, which for long has been known to be in the range
$1-10$~\cite{kolbturner90}.  Thus by observing just one primordial
light element one can predict the abundances of all the other light
elements.

The deuterium to hydrogen abundance, D/H, is observed by Ly-$\alpha$
features in several quasar absorption systems at high red-shift, $D/H
= 2.78^{+0.44}_{-0.38}\times10^{-5}$~\cite{kirkman2003}, which in SBBN
translates into the baryon abundance, $\eta_{10} = 5.9\pm0.5$.  Using
SBBN one thus predicts the helium mass fraction to be in the range $0.2470
< Y_p < 0.2487$.  The dispersion in various deuterium observations
is, however, still rather large, ranging from $D/H = 1.65\pm
0.35\times10^{-5}$~\cite{pettini2001} to $D/H =
3.98^{+0.59}_{-0.67}\times10^{-5}$~\cite{kirkman2003}, which most
probably indicates underestimated systematic errors.

\begin{figure}[htb]
\begin{center}
\includegraphics[angle=-90, width=3.in]{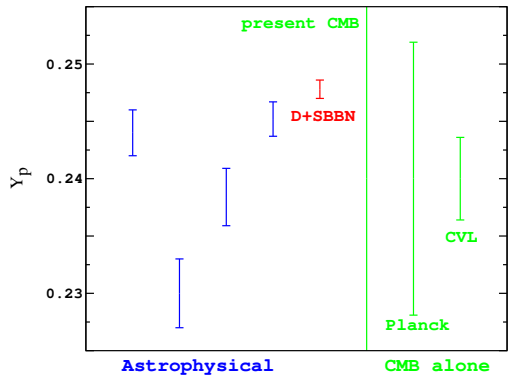}
\end{center}
\caption{On the left (blue) we plot a few current direct astrophysical
measurements of the helium mass
fraction $Y_p$ with their $1-\sigma$ statistical
errors, and the
value inferred from deuterium measurements combined with SBBN (red) (see text
for references). On the
right (green), a direct comparison
with CMB present-day accuracy (actual data, this work; the errorbar
extends in the range $0.16 < Y_p < 0.50$) and with its future potential
(Fisher matrix forecast for Planck and a Cosmic Variance Limited experiment).}
\label{fig:compilation}
\end{figure}

The observed helium mass fraction comes from the study of
extragalactic HII regions in blue compact galaxies.  One careful
study~\cite{IzotovThuan} gives the value $Y_P = 0.244 {\pm} 0.002$;
however, also here there is a large scatter in the various observed
values, ranging from $Y_p = 0.230 \pm 0.003$~\cite{oliveSS1997} over
$Y_p = 0.2384 \pm 0.0025$~\cite{peimbert2002} and $Y_p = 0.2391 \pm
0.0020$~\cite{luri03} to $Y_p=0.2452\pm0.0015$~\cite{IT1999}.  Besides
the large scatter there is also the problem that the helium mass
fraction predicted from observation of deuterium combined with SBBN,
$0.2470 < Y_p < 0.2487$, is larger than (and seems almost in
disagreement with) most of the observed helium abundances, which
probably points towards underestimated systematic errors, rather than
the need for new physics~\cite{olive03,Barger:2003zg}.
Figure~\ref{fig:compilation} is a compilation of the above
measurements, and offers a direct comparison with the current (large)
errors from CMB observations (presented in section III below) and with
the potential of future CMB measurements (discussed in section III D).

The observed abundance of primordial $^7\textrm{Li}$ using the Spite plateau is
possibly spoiled by various systematic effects~\cite{Ryan,Salaris}.
Therefore it is more appropriate to use the SBBN predictions
together with observations to estimate the depletion factor
$f_7=^7\textrm{Li}_{\textrm{obs}}/^7\textrm{Li}_{\textrm{prim}}$
instead of using $^7\textrm{Li}_{\textrm{obs}}$ to infer the value of
$\eta$~\cite{burlesNT01,hansen02,steig99,steig02}.

The numerical predictions of standard BBN (as well as various
non-standard scenarios) have reached a high level of
accuracy~\cite{hata95,burlesNT01,lopezturner1999,Emmp1,Emmp2,Cuoco:2003cu}, and the
precision of these codes is well beyond the systematic errors
discussed above.

\subsection{The role of neutrinos}

If the CMB-determined helium mass fraction turns out to be as high as
suggested by SBBN calculations together with the CMB observation of
$\Omega_bh^2$ (as discussed above), this could indicate a systematic
error in the present direct astrophysical helium observations.
 Alternatively, if the CMB could independently determine the helium
 value with sufficient precision to confirm the present helium
observations, then this would be a smoking gun for new physics. In
fact, one could easily imagine non-standard BBN scenarios which would
agree with present observations of $\eta_{10}$, while having a low
helium mass fraction.  All that is needed is additional
non-equilibrium electron neutrinos produced at the time of neutrino
decoupling which would alter the $n-p$ reaction. This could alter the
resulting helium mass fraction while leaving the deuterium abundance
unchanged. One such possibility would be a heavy sterile neutrino
whose decay products include $\nu_e$. A sterile neutrino with
life-time of $1-5$ sec and with decay channel $\nu_s \rightarrow \nu_e
+ \phi$ with $\phi$ a light scalar (like a majoron), would leave the
deuterium abundance roughly untouched, but can change the helium mass
fraction between $\Delta Y_p = -0.025$ and $\Delta Y_p = 0.015$ if the
sterile neutrino mass is in the range $1-20$ MeV~\cite{dolgov98}. A
simpler model would be standard neutrino oscillation between a sterile
neutrino and the electron neutrino. The lifetime is about 1 sec when
the sterile state has mass about 10 MeV, and the decay channel is
$\nu_s \rightarrow \nu_e +l + \bar l$ (with $l$ any light lepton), and
such masses and life-times are still unconstrained for large mixing
angle~\cite{dolgov00} (related BBN issues are discussed in
refs.~\cite{dibari,kirilova,abaz,dolvil}).  Such possibilities are hard to
constrain without an independent measurement of the helium mass
fraction.

Another much studied effect of neutrinos is the increased
expansion rate of the universe if additional degrees of freedom are
present (for BBN), and the degeneracy between the total density in
matter and relativistic particles (for CMB).  This issue has recently
been studied in detail in refs.~\cite{recentnu,Barger:2003zg}
in view of the new
WMAP results, and we need not discuss this further here.
We thus fix $N_\nu = 3.04$ \cite{3neutrinos}.
Also an electron neutrino chemical potential could potentially
alter the BBN predictions~\cite{kangsteigman}, however, with
the observed neutrino oscillation parameters the different neutrino
chemical potentials would equilibrate before the onset of
BBN~\cite{nuosc},
hence virtually excluding this possibility (see however~\cite{dannygary}).

\section{Cosmic Microwave Background}
\label{sec:cmb}

\subsection{Photon recombination and reionization}

The recent WMAP data allow one to determine with very high
precision the epoch of photon decoupling, $z_{\rm dec}$, \ie the epoch at
which the ionized electron fraction, $x_e(z) = n_e/n_H$, has
dropped from 1 to its residual value of order $10^{-4}$. Here
$n_e$ denotes the number density of free electrons, while
$n_H$ is the total number density of H atoms (both ionized and
recombined). After this moment, photons are no longer coupled to electrons (last
scattering), and they free stream. The redshift of decoupling has been
determined to be $z_{\rm dec}=1088^{+1}_{-2}$ \cite{spergel03}, which
corresponds to a temperature of about 0.25 eV. Helium
recombines earlier than hydrogen, roughly in two steps:  around
redshift $z=6000$ HeIII recombines to HeII, while HeII to HeI
recombination begins around $z<2500$ and finishes just after the
start of H recombination~(see e.g.~\cite{lyubarsky83,hu95,seager99, seager00}).

\begin{figure}[thb]
\begin{center}
\includegraphics[width=3.in, angle=-90]{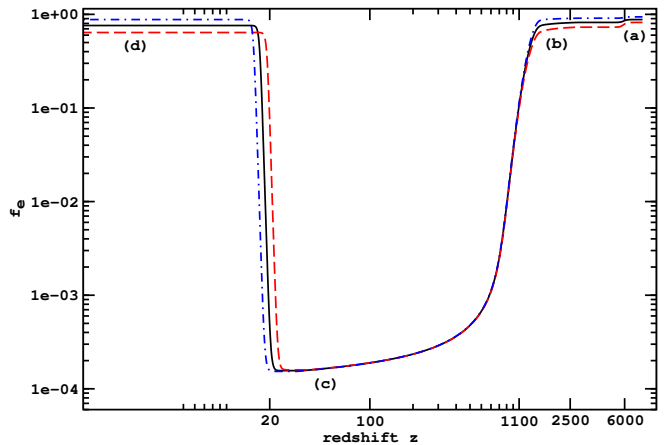}
\end{center}
\caption{Evolution of the number density of electrons
normalized to the number density of baryons, $f_e=n_e/n_b$,
as a function of redshift for different values of the helium
fraction $Y_p$. The black-solid curve corresponds to the standard value
$Y_p = 0.24$, the red-dashed (blue-dot-dashed) to $Y_p=0.36$ ($Y_p=0.12$). The
labels (a) to (d) indicate the four different phases discussed
in the text.}
\label{f:ionhist}
\end{figure}

Denoting by $n_{He}$ and $n_b$ the number densities per m$^3$
of He atoms and baryons, respectively, the helium mass fraction
is defined as $Y_p = 4 n_{He}/n_b$. The baryon number density is
related to the baryon energy density today, $\omega_b$, by
$n_b = 11.3 (1+z)^3 \omega_b$ and we have $n_H = n_b(1-Y_p)$.
Usually, the ionization history is described in terms of
$x_e(z) = n_e/(n_b(1-Y_p))$. However, for the purpose of discussing
the role of $Y_p$, it is
more convenient to consider the quantity $f_e(z) = n_e/n_b$ instead, the
ratio of free electrons to the total number of baryons. For brevity,
we will call $f_e$ the free electron fraction.
Once the
baryon number density has been set by fixing $\omega_b$, one can
think of $Y_p$ as an additional parameter which controls the number
of free electrons available in the tight coupling regime.
The CMB power spectrum depends on the full detailed evolution of the
free electron fraction, but we can qualitatively describe the role
of helium in four different phases of the ionization/recombination
history (see Fig.~\ref{f:ionhist}).

\begin{itemize}
\item[(a)] Before HeIII recombination all electrons are free, therefore
$f_e(z>6000) = 1 - Y_p/2$.
\item[(b)] HeII progressively recombines and just before
H recombination begins, $f_e$ has dropped to the value
$f_e(z \approx 1100) = 1 - Y_p$.
\item[(c)] After decoupling, a residual fraction of free electrons freezes out,
giving $f_e(30 \lsim z \lsim 800) = \fres \approx
2.7 \cdot 10^{-5} \sqrt{\omega_m}/\omega_b$.
\item[(d)] Reionization of all the H atoms gives
$f_e(z \lsim 20) = 1 - Y_p$.
\end{itemize}

During phase (a), the photon-baryons fluid is in the tight coupling
regime. However the presence of ionized He increases diffusion damping,
therefore having an impact on the damping scale in the acoustic peaks
region.
When the detailed energy levels structure of HeII is taken into
account~\cite{seager99}, the transition to phase (b) is smoother than
in the Saha equation approximation.  Therefore the plateau with
$f_e = 1 - Y_p$ is not visible in Fig.~\ref{f:ionhist}. Before H recombination,
He atoms remain tightly coupled to H atoms through collisions,
with the same dynamical behaviour. In particular, it is the total
$\omega_b$ which determines the amount of gravitational pressure on
the
photon-baryons fluid, and which sets the acoustic peak enhancement/suppression. Hence
we do not expect the value of $Y_p$ to have any influence on the boosting
(suppression) of odd (even) peaks.
The redshift of decoupling (transition between (b) and (c)) depends mildly
on $Y_p$ in a correlated way with $\omega_b$, since the number
density of free electrons in the tight coupling regime (just before H
recombination) scales as
$n_e = f_e n_b = n_b(1-Y_p)$. Hence an increase in $\omega_b$ can be compensated
by allowing for a larger helium fraction.
An analytical estimate along the same lines
as in e.g.~\cite{kolbturner90} indicates
that a $10\%$ change in $Y_p$ affects $\zdec$ by roughly $0.1\%$,
which corresponds to $\Delta \zdec \approx 1$. This is of the same
order as the current 1-$\sigma$ errors on $\zdec$, obtained by fixing
$Y_p = 0.24$.

After H recombination, the residual ionized electron fraction $\fres$ does
not depend on $Y_p$, but is inversely proportional to the total baryon density
(phase (c)).
As the CMB photons propagate, they are occasionally rescattered by
the residual free electrons. The corresponding optical depth,
$\tres$ is given by
\begin{eqnarray}
\tres & =& \int_{t_0}^{t_{\textrm{dec}}}n_e^{\textrm{res}} c \sigma_T dt \\
      & \approx& 1.86\cdot 10^{-6}
        \int_0^{\zdec} \frac{(1+z)^2}{((1+z)^3 + \OLa/\Omega_m)^{1/2}} dz \nonumber .
\end{eqnarray}
Performing the integral we can safely neglect the contribution of
the cosmological constant at small redshift, since $\zdec \gg
\OLa/\Omega_m$. Retaining only the leading term, the approximated
optical depth from the residual ionization fraction is estimated
to be
\begin{equation}
\tres \approx 1.24\cdot10^{-6} (1+\zdec)^{3/2} \approx 0.045,
\end{equation}
{\it independent} of the cosmological parameters and of the helium fraction.
Therefore after last scattering we do not expect any significant effect
on CMB anisotropies coming from the primordial helium fraction,
until the reionization epoch.

Fairly little is known about the exact reionization mechanism and
its redshift dependence (for a review see e.g.~\cite{haiman}).
Observation of Gunn-Peterson troughs indicate that the universe
was completely ionized after redshift $z \approx 6$, when the
universe seemingly completed the reionization~\cite{BeckerFan},
possibly for the second time~\cite{cen}. If temperature
information only is available, CMB anisotropies are sensitive only
to the integrated reionized fraction, represented by the optical
depth, independent of the specific reionization history. However,
specific signatures are imprinted on the E-polarization and
ET-cross correlation power spectra by the detailed shape of the
reionization history (for a detailed discussion, see \cite{Bruscoli02, Hu03,
K03, Holder03}). There are several physically motivated
reionization scenarios, which however cannot be clearly
distinguished at present \cite{Haiman03,Hansen:2003yj}.
In this work we use the
most simple model, the sudden reionization scenario: we assume
that at the reionization redshift $\zreion$ all the hydrogen was
quickly reionized, thus producing a sharp rise of $n_e$ from its
residual value to $n_H$. More precisely, $\zreion$ is the redshift
at which $x_e(\zreion) = 0.5$. In our treatment we neglect HeII
reionization, for which there is evidence at a redshift $z \approx
3$ (see \cite{frieman02} and references therein). This effect is
small, since the extra electron released at $z \approx 3$ would
change the reionization optical depth by about only $1\%$. We also
neglect the increase of the helium fraction due to non-primordial
helium production, which has a negligible effect on CMB
anisotropies. Those approximations do not affect the results at
today's level of sensitivity of CMB data: for WMAP noise levels,
even inclusion of the polarization spectra is not enough to
distinguish between a sudden reionization scenario and a more
complex reionization history. At the level of Planck a more
refined modelling of the reionization mechanism will be
necessary~\cite{Holder03,naselsky03}.

In the sudden reionization scenario adopted here, the relation
between reionization redshift and reionization optical depth,
$\treion$, is given by
\begin{eqnarray}\label{eq:zreion}
\treion & = & \int_{t_0}^{t_{\textrm{reion}}}n_e c \sigma_T dt \nonumber \\*
        & \approx & 11.3 c \sigma_T \omega_b (1-Y_p) \int_0^{\zreion}
      \frac{d\eta}{da} dz ,
\end{eqnarray}
where  $t$ is physical time, $\eta$ is conformal time and $a$ the scale factor.
Here again, since the number density of reionized electrons scales as
$\omega_b(1-Y_p)$, the redshift of reionization is positively correlated
with $Y_p$ (for fixed optical depth and baryon density).

\begin{figure}[thb]
\begin{center}
\includegraphics[width=3.in]{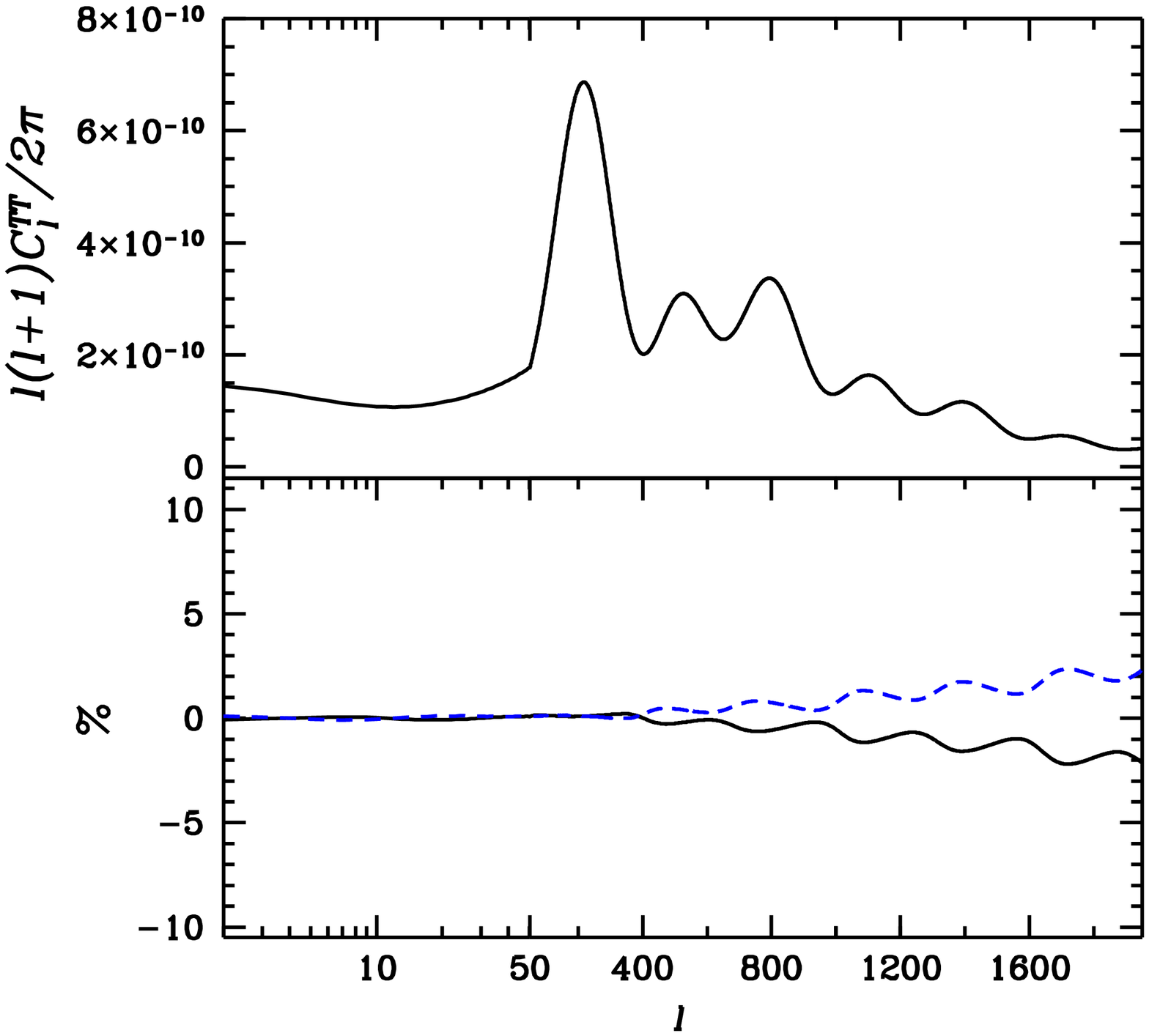}
\includegraphics[width=3.in]{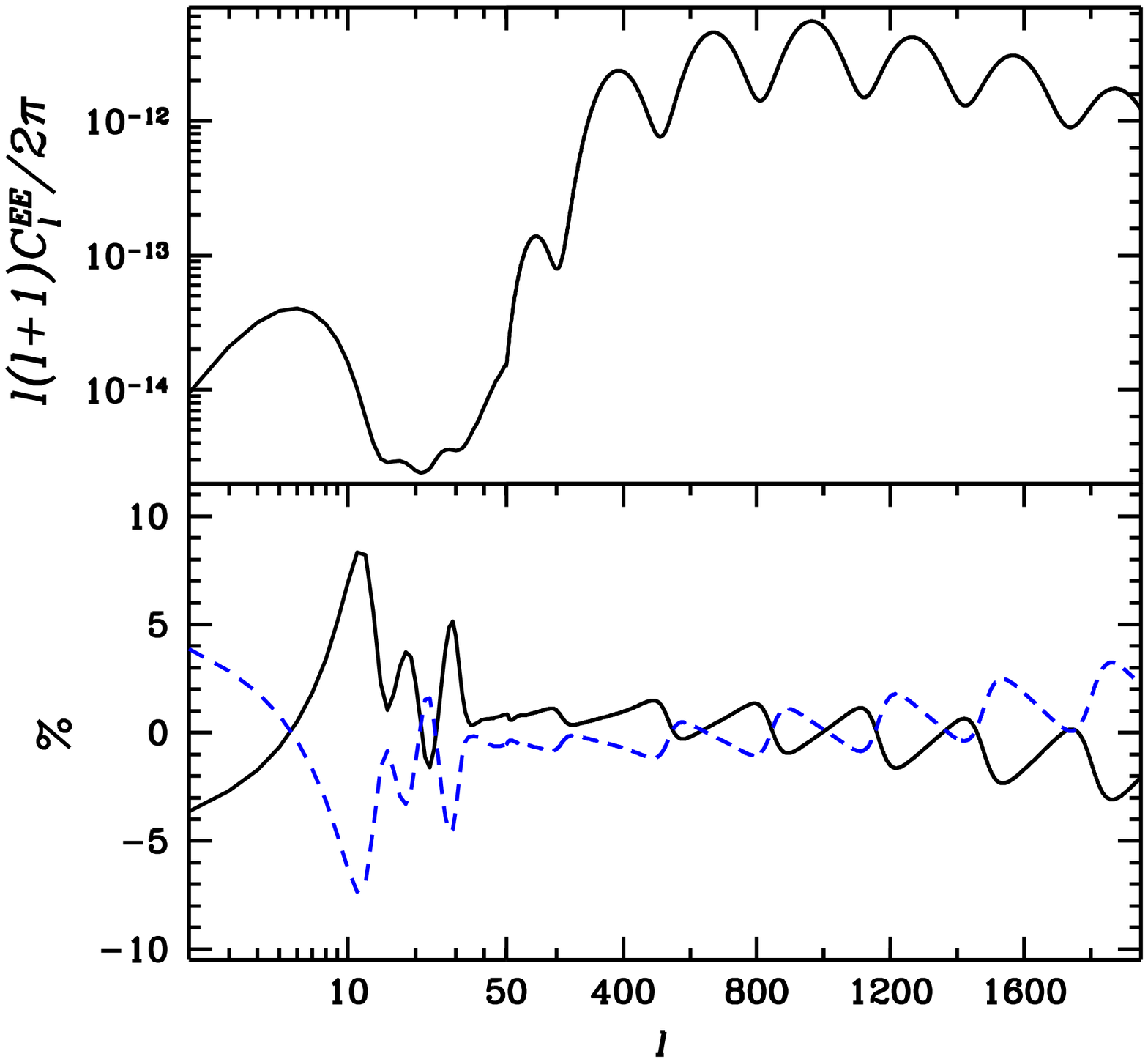}
\end{center}
\caption{CMB temperature (top panel) and polarization
  (bottom panel) power spectra and percentage difference with two
  different values of the helium fraction for a standard $\Lambda$CDM
  model. The solid-black (dashed-blue) line corresponds to a 10\% larger
  (smaller) value of $Y_p$ wrt to the standard value, $Y_p = 0.24$.
  All other parameters are fixed to the value of our fiducial model
  (Table~\ref{t:fiducial}),
  in particular, we have $\treion=0.166$.}
\label{fig:effect}
\end{figure}

As a result of the physical mechanism described above, a
$10\%$ change in $Y_p$ has a net impact on the CMB power spectrum at the
percent level.  The impact on the CMB temperature and
polarization power spectra is highlighted in Fig.~\ref{fig:effect}.
In the temperature panel, we notice that a larger helium fraction
slightly suppresses the peaks because of diffusion damping, while
it has no impact on large scales. Polarization is induced by the
temperature quadrupole component at last scattering.  When
reionization occurs, there is a generation of polarized power on
the scale corresponding to the acoustic horizon size at the
reionization redshift. This particular signature is called the
``reionization bump'', and is clearly visible in the bottom panel
of Fig.~\ref{fig:effect} in the $\ell \approx 10$ region. The
position of the bump in multipole space scales as
$\ell_{\textrm{bump}} \propto
\sqrt{\zreion}$~\cite{Zaldarriaga97}. As discussed above, a change
in the helium fraction implies a shift of the redshift of
reionization for a given (fixed) optical depth, Therefore the
value of  $Y_p$ has an effect on the position of the reionization
bump in the polarization power spectrum, but not on its height,
which is controlled by the optical depth and is proportional to
$\tau^2$. This effect is highlighted in the bottom panel: a 10\%
change in $Y_p$ induces roughly a 10\% change in the position of
the bump. The subsequent two oscillatory features for $\ell \lsim
50$ reflect the displacement of further secondary, reionization
induced polarization oscillations. However, since the value of
polarized power is very low in that region, such secondary
oscillations are very hard to detect precisely. In principle,
given an accurate knowledge of the reionization history, the
effect of $Y_p$ on the polarization bump would assist into
determining the helium abundance. However, our ignorance of the
reionization history prevents us from recovering useful
information out of the measured reionization bump. The
displacement induced by $Y_p$ is in fact degenerate with a partial
reionization, or with other, more complex reionization mechanisms
(see~\cite{Holder03}). Hence constraints on $Y_p$ come effectively
from the damping tail in the $\ell \gsim 400$ region of the
temperature spectrum, which needs to be measured with very high
accuracy.

Other light elements like deuterium and helium-3 are much
less abundant, and will therefore have even smaller effect on the CMB
power spectrum, at the order of $10^{-5}$.

\subsection{Monte Carlo analysis}

We use a modified version of the publicly available Markov Chain Monte
Carlo package \textsc{cosmomc}~\cite{cosmomc} as described in
\cite{lewis02} in order to construct Markov Chains in our 7
dimensional parameter space.  We sample over the following set of
cosmological parameters: the physical baryon and CDM densities,
$\omega_b \equiv \Omega_b h^2$ and $\omega_c \equiv \Omega_c h^2$, the
cosmological constant in units of the critical density,
$\Omega_{\Lambda}$, the scalar spectral index and the overall
normalization of the power spectrum, $n_s$ and $A_s$ (see section III D
below for a more precise definition), the redshift at
which the reionization fraction is a half, $\zreion$, and the
primordial helium mass fraction, $Y_p$.  We restrict our analysis to
flat models, therefore the Hubble parameter, $h = H_0/100 \text{ km}
\text{ s}^{-1} \text{Mpc}^{-1}$, is a derived parameter, $h =
[(\omega_c + \omega_b)/(1-\Omega_{\Lambda})]^{1/2}$.  We consider
purely adiabatic initial conditions, and we do not include
gravitational waves. In the CMB analysis, we assume 3 massless
neutrino families and no massive neutrinos.  We include the WMAP data
\cite{kogut03, hinshaw03} (temperature and polarization) with the
routine for computing the likelihood supplied by the WMAP team
\cite{verde03}. We make use of the CBI \cite{pearson02} and of the
decorrelated ACBAR~\cite{acbar,kuo03} band powers above $\ell = 800$
to cover the small angular scale region of the power spectrum.

Since $Y_p$ is a rather flat direction in parameter space with
present-day data, we find that a much larger number of samples is
needed in order to achieve good mixing and convergence of the chains
in the full 7D space. We use $M=4$ chains, each containing
approximately $N=3 \cdot 10^{5}$ samples. The mixing diagnostic is
done on the same lines as in \cite{verde03}, by means of the Gelman
and Rubin criterion \cite{gelman92}.  The burn-in of the chains also
takes longer than in the case where $Y_p$ is held fixed, and we
discard 6000  samples per chain.

\subsection{CMB analysis results}
\label{sec:results}

Marginalizing over all other parameters, we find that the helium mass
fraction from CMB alone is constrained to be $Y_p < 0.647$ at 99\%
c.l. (1 tail limit), and
\begin{equation}
0.160 < Y_p < 0.501
\end{equation}
at 68\% c.l. (2 tails).
Thus, for the first time the primordial helium mass fraction has been
observed using the cosmic microwave background. However, present-day
CMB data do not have sufficient resolution to discriminate between the
astrophysical helium measurements, $Y_p \sim 0.244$, and the deuterium
guided BBN predictions, $Y_p \sim 0.248$, which would require
percent precision.

\begin{figure}[htb]
\begin{center}
\includegraphics[width=3.in]{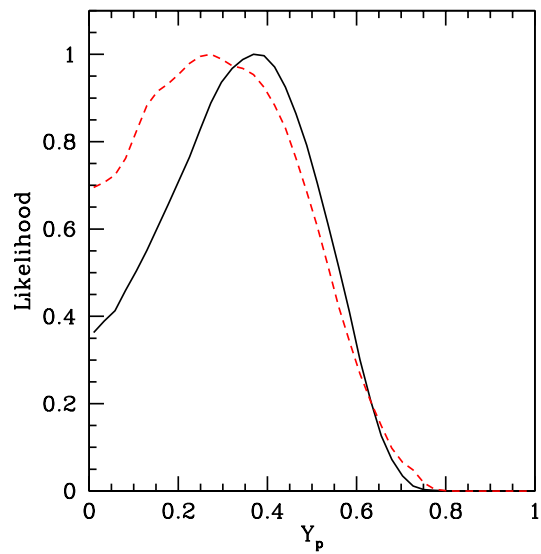}
\end{center}
\caption{One-dimensional posterior likelihood distribution for the helium mass
fraction, $Y_p$, using CMB data only. The solid-black line
is for all other parameters marginalized, the dashed-red line gives
the mean likelihood.}
\label{fig:Ypmarg}
\end{figure}

In Figure \ref{fig:Ypmarg} we plot the marginalized and the mean
likelihood of the Monte Carlo samples as a function of $Y_p$.  If the
likelihood distribution is Gaussian, then the 2 curves should be
indistinguishable. The difference between marginalized and
mean likelihood for $Y_p$ indicates that the marginalized parameters
are skewing the distribution, and therefore that correlations play
an important role. Although the
mean of the 1D marginalized likelihood is rather high,
$\langle \mathcal{L}(Y_p) \rangle =
0.33$, the mean likelihood peaks in the region indicated by
astrophysical
measurements, $Y_p \sim 0.25$. In view of this difference, it is
important to understand the role of correlations with other
parameters, and we will turn to this issue now.

\begin{figure}[thb]
\begin{center}
\includegraphics[width=3.in]{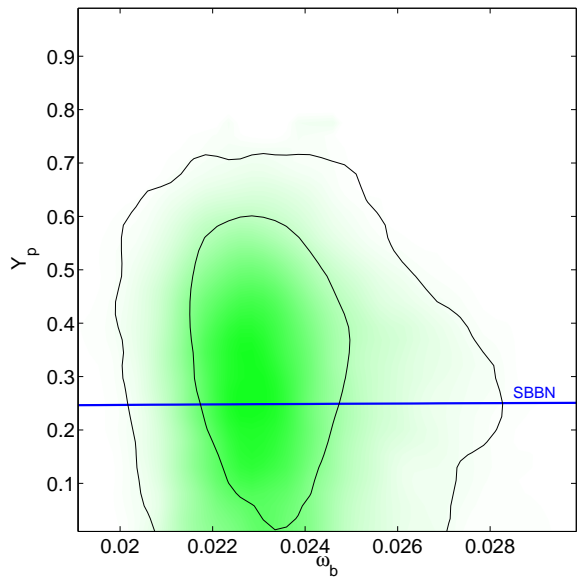}
\end{center}
\caption{Joint 68\% and 99\% confidence
contours in the ($\omega_b,
Y_p$)-plane from CMB data alone. The solid-blue line gives
the SBBN prediction~\cite{burlesNT01}, which on this figure almost looks
like a straight line.}
\label{fig:obyp}
\end{figure}

In Figure \ref{fig:obyp} we plot joint 68\% and 99\% confidence
contours in the ($\omega_b, Y_p$)-space.
From the Monte Carlo samples we obtain a small and negative
correlation coefficient between the two parameters $\text{corr}(Y_p,
\ob )= -0.14$. Baryons and helium appear to be
anticorrelated simply because present-day WMAP data do not map
the peaks structure to sufficiently high $\ell$. Precise measurements
in the small angular scale region should reveal the expected
positive correlation between the baryon and helium abundances, which
is potentially important in order to correctly combine BBN
predictions and CMB measurements of the baryon abundance. We turn to
this question in more detail in the next section.
In SBBN the baryon fraction and helium fraction are correlated along
a different direction (cf.~Fig.~\ref{fig:obyp}). However,
this correlation is very weak, and the SBBN relation gives practically a flat
line.
Since the
two parameters are not independent from the CMB point of view, it is
in fact not completely accurate to do the CMB analysis with fixed helium mass
fraction of $Y_p = 0.24$ to get the error-bars on the baryon fraction,
and then re-input this baryon fraction (and error-bars) to predict the
helium mass fraction from BBN. The most accurate procedure is
to analyse the CMB data leaving $Y_p$ as a free parameter, thereby
obtaining the correct (potentially larger) error-bars on $\omega_b$
upon marginalization over $Y_p$.

In view of the emerging baryon tension between CMB and BBN, it is
important to check whether allowing helium as a free parameter  can significantly
change the CMB determination of the baryon density or its error.
In order to evaluate in detail the impact of $Y_p$ on the error-bars for
$\omega_b$, we consider the following 3 cases.
\begin{itemize}
\item[(a)] The usual case, when the
helium fraction for the CMB analysis is assumed to be known
{\it a priori} and is fixed to the canonical value $Y_p=0.24$.
\item[(b)] A  case with a
weak astrophysical Gaussian prior on the helium fraction, which we
take to be $Y_p = 0.24 \pm 0.01$. As discussed above, the error-bars
of the astrophysical measurements are typically a factor 5 tighter than this, but our
prior is chosen to encompass the systematic spread between
the different observations.
\item[(c)] The case in which we assume a uniform prior for $Y_p$ in
  the range $0 \leq Y_p \leq 1$, \ie $Y_p$ is considered as a totally
  free parameter.
\end{itemize}

We do not find any significant change in the error-bars for $\omega_b$
in the 3 different cases. The confidence intervals on $\ob$ alone are
determined to be (case (c))  $0.0221 < \omega_b < 0.0245$ at 68\%
c.l. ($0.0204 < \omega_b < 0.0276$ at 99 \% c.l.). The standard deviation
of $\omega_b$ as estimated from the Monte Carlo samples is found to be
$\hat{\sigma}_b = 1.3 \cdot 10^{-3}$. This is in complete agreement
with the error-bars on $\omega_b$ obtained by the WMAP team for the
standard $\Lambda$CDM case \cite{spergel03}. We conclude that at the
level of precision of present-day CMB data, it is still safe to treat
the baryon abundance and the helium mass fraction as independent
parameters. This result is non-trivial, since the fact that the damping
tail is not yet precisely measured above the second peak would a priori suggest that
degeneracies between $Y_p$, $\omega_b$ and $n_s$ could potentially play a
role once the assumption of zero uncertainity on $Y_p$ is relaxed. The impact
of $Y_p$ is small enough, and the error-bars on $\omega_b$ large enough
that a uniform prior on $Y_p$ can still be accomodated within the uncertainity
in the baryon abundance obtained for case (a).
However, the $Y_p-\omega_b$ correlation will have to be
taken into account to correctly analyse future CMB data, with a
quality such as Planck.  We discuss this potential in the next
section.

\begin{figure}[htb]
\begin{center}
\includegraphics[width=3.5in]{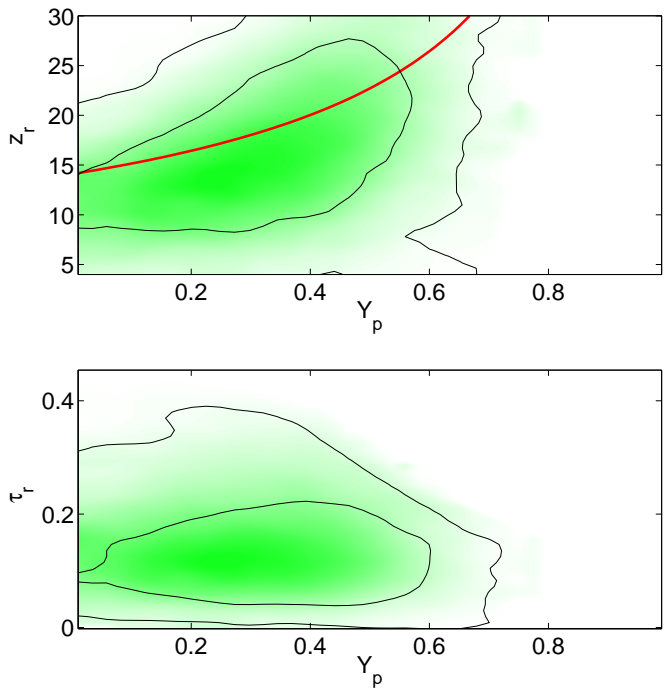}
\end{center}
\caption{Joint 68\% and 99\% confidence
  contours in the ($Y_p, \zreion$)-plane (upper panel) and in the
  corresponding ($Y_p, \treion$)-plane (bottom panel) from CMB data
  alone.
  In the upper panel, the
  solid-red line is the relation $\zreion(Y_p)$ from eq. (\ref{eq:zreion}),
  obtained by fixing the reionization optical depth to the value $\treion = 0.166$,
  while the other parameters are the ones of our fiducial $\Lambda$CDM
  model.  Although clearly the exact shape of $\zreion(Y_p)$ depends
  on the particular choice of cosmology, it is apparent that the
  $Y_p-\zreion$ degeneracy is along this direction. The correlation
  between $Y_p-\treion$ is almost negligible with present-day data
  (bottom panel).}
\label{fig:ypzr}
\end{figure}

We observe the expected correlation between the redshift of reionization
and the helium
fraction (Fig.~\ref{fig:ypzr}), which is discussed above.
The correlation coefficient between the two parameters is found to be
rather large and positive, $\text{corr}(Y_p, \zreion )=0.40$.
This correlation produces a noticeable change in the
marginalized 1D-likelihood distribution for $\zreion$ as we go from
case (a) to case (c). Marginalization over the additional degree of
freedom given by $Y_p$ broadens considerably the error-bars on
$\zreion$.  In fact, the 68\% confidence interval for $\zreion$
increases by roughly 20\% (and shifts  to somewhat higher values),
from $10.2-20.9$ (case (a)) to $10.6-23.3$ (case (c)).  Case (b)
exhibits similar error-bars as case (a). On the other hand, the
determination
of the reionization optical depth is not affected by the inclusion of
helium as a free parameter, giving in all cases $0.08 < \treion <
0.23$.
Correspondingly, the correlation is less
significant, $\text{corr}(Y_p, \treion )=-0.11$.
We therefore conclude that the differences in the determination of
$\zreion$ are due only to the variation of the amount of electrons
available for reionization as $Y_p$ is changed.

\begin{figure}[htb]
\begin{center}
\includegraphics[width=3.5in]{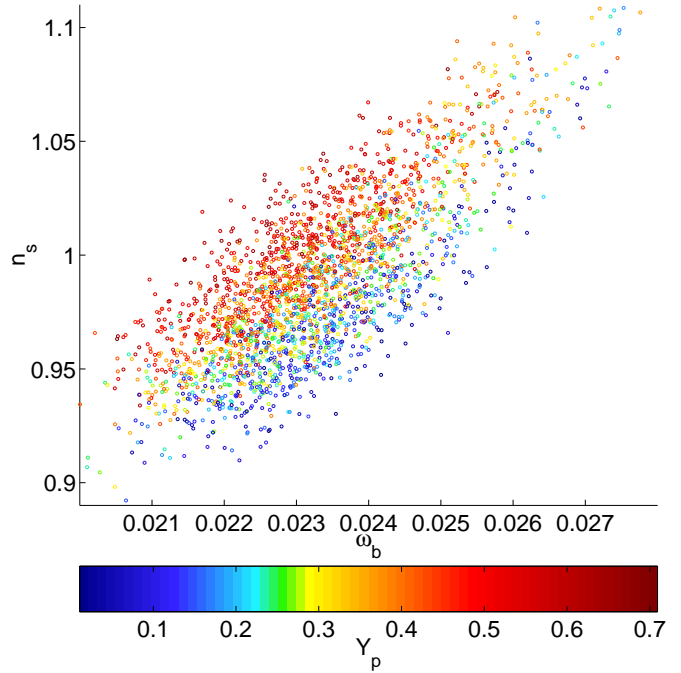}
\end{center}
\caption{Scatter plot in the $\ob-n_s$ plane, with the value of
  $Y_p$ rendered following the colour scale. Green corresponds roughly
  to the SBBN preferred value.}
\label{fig:3D}
\end{figure}

Leaving $Y_p$ as a free parameter also has an impact on the relation
between $\ob$ and the scalar spectral index, $n_s$. The extra power
suppression on small scales which is produced by a larger $Y_p$ can
be compensated by a blue spectral index (see
Fig.~\ref{fig:3D}).

\subsection{Potential of future CMB observations}
\label{sec:future}

In order to estimate the precision with which future satellite
CMB measurements  will be able to constrain the helium mass fraction
we perform a Fisher matrix analysis (FMA). This technique approximates
the likelihood function with a Gaussian distribution around a fiducial
model, which is assumed to be the best fit model. The Fisher
information matrix $F$ gives the second order expansion of the
likelihood around its peak, and it is computed from the derivatives of the power spectrum
with respect to the cosmological parameters.
The expected performance of the
experiment can be modelled with a noise contribution to the likelihood
function, which is described in terms of a few experimental parameters.
The covariance matrix $C$ is then given by the
inverse of the Fisher matrix, $C=F^{-1}$. It is then straightforward to evaluate the
expected 1-$\sigma$ error on parameter $i$, which is given by $\sqrt{c_{ii}}$
(all other marginalized).
The main advantage of the FMA is that it gives reliable and accurate predictions
(including information on the expected degeneracies) with minimal computational
effort.
For further details on the Fisher Matrix formalism,
see e.g.~\cite{knox95,bond97,zaldarriaga97, tegmark96,
  EB99,bowen,alpha1, alpha2}.

\subsubsection{Parameters set}

In order to obtain a reliable prediction, it is extremely important to
choose a parameter set wrt which the dependence of the CMB power spectrum
is as linear and uncorrelated as possible.
This issue has been discussed exhaustively in
Ref.~\cite{kosowsky02}, where the authors introduce a set of ``physical parameters''
which satisfies the above requirements. In the present work we retain most of
the physical parameters defined in ref.~\cite{kosowsky02}: the ratio between the
sound horizon at decoupling and the angular diameter distance $\kA$,
the baryon density $\kB = \Omega_bh^2$, the energy density in the cosmological
constant $\kV=\Omega_{\Lambda}h^2$, the matter-radiation density at decoupling
$\kR$ and $\kM$, which is mainly a function of the matter and radiation content.
We adopt a slightly different choice for the physical
parameter describing reionization. For adiabatic perturbations, the
initial power spectrum of the gauge invariant curvature perturbation
$\zeta$ is written as
$$
P_\zeta(k) = A_s \left( \frac{k}{k_0} \right) ^ {n_s -1}
$$
(and we do not take running into account). The quantity $\zeta$
corresponds to the intrinsic curvature perturbation on comoving
hypersurfaces, and at the end of inflation is related to the
gravitational potential perturbation, $\Psi$, by $\zeta =
\frac{3}{2}\Psi$ (see e.g.~\cite{durrer} for more details). We take
the pivot-scale $k_0$ to be $k_0 = 0.05\text{ Mpc}^{-1}$.
If $\treion$ denotes the optical depth to reionization,
then defining $\rZ = A_s \exp(-2 \tau)$ is a good way to take into account
the degeneracy between the optical depth and normalization. Our parameter set
contains then the six above physical parameters ($\kA, \kB, \kV, \kR,
\kM, \rZ$), the power spectrum normalization $A_s$,
the scalar spectral index $n_s$ and the helium fraction $Y_p$.

The choice of the physical parameter set makes it easy to implement
in the FMA interesting theoretical priors. For instance, we are interested
in imposing flatness in our forecast, in order to be able to directly compare
present-day accuracy on $Y_p$ with the potential of Planck and CVL. The prior
on the curvature of the universe is imposed in the FMA by
fixing the value of the parameter $\kA$ to the one of the fiducial model.
In fact, the parameter  $\kA$ is a generalization of the shift parameter, which
describes the sideway shift of the acoustic peak structure of the CMB power
spectrum as a function of the geometry of the universe and its content in matter, radiation
and cosmological constant. Although imposing $\kA=$ const is not
the same as having curvature=constant over the full range of cosmological parameters,
for the purpose of evaluating derivatives the two conditions reduce to the same.
The fact that our fiducial model is actually slightly open (see
below),
does not make any substantial
difference in the results, apart from reducing the numerical inaccuracies which would
arise had we computed the derivatives around an exactly flat model.
We can also easily impose a prior knowledge of the helium fraction,
by fixing the value of $Y_p$, as it is usually done in present CMB
analysis, and investigate
how this modifies the expected error on the the baryon density.

\subsubsection{Accuracy issues}

We numerically
compute double sided derivative of the power spectrum around the
fiducial model with cosmological parameters given in Table~\ref{t:fiducial}.
We find it necessary to increase the accuracy of CAMB by a factor of 3 in each of
the ``accuracy boost'' values. As a fiducial model, we use the best fit model to the
WMAP data
for the standard $\Lambda$CDM scenario, as given in Table 1 of ref.~\cite{spergel03}.
However, in order to avoid numerical inaccuracies which arise when differentiating around
a flat
model, we reduce slightly the value of $\Omega_\Lambda$ by imposing an open universe,
$\Omega_{\textrm{tot}} = 0.99$.
 We perform the FMA for the expected capabilities of Planck's High
Frequency Instrument (HFI) and for an ideal CMB measurement which would be cosmic variance
limited (CVL) both in temperature and in E-polarization (and we do not consider the
B-polarization spectrum), and therefore represents the best possible parameter measurement
from CMB anisotropies alone. The complicated issues coming from foreground removals,
point source subtractions, etc. are assumed to be already (roughly) taken into account in
the experimental parameters for the experiment. Those are the effective percentual
sky coverage $\fsky$, the number of channels, the
sensitivity of each channel $\sigma_c^{T,E}$ for temperature (T) and E-polarization
(E) in $\mu$K and the angular resolution $\theta_c^{T,E}$
(in arcmin). For Planck HFI, we take the 3 channels with frequencies
100, 143 and 217 GHz, with respectively
$\sigma_{c=1,2,3}^{T}= 5.4, 6.0, 13.1$ and
$\sigma_{c=2,3}^{E}= 11.4, 26.7$ and we have $\fsky=0.85$ \cite{PlanckHP}
Since the CVL is an ideal experiment, we put its noise to zero and assume perfect
foregrounds removal, so that   $\fsky=1$. In order to test the accuracy of our
predictions and compare present-day results with the forecasts, we also perform an FMA
with WMAP first year parameters,
obtaining excellent agreement between the FMA results and the error-bars from
actual data. For the purpose of comparison, we include forecasts for the full WMAP
4 years mission, which will also measure E-polarization and reduce present-day errors
on the temperature spectrum by a factor of 2. We limit the range of multipoles to
$\ell < 2000$, because at smaller angular scales non-primary anisotropies begin to
dominate (Sunyaev-Zeldovich effect). The authors of ref.~\cite{precision03} discuss the issue of
numerical precision of 3 different CMB codes and conclude that they are accurate to within
$0.1\%$. While this is encouraging, it is not of direct relevance to this work, since
what matters in the computation of derivatives is not much the absolute precision of the
spectra, but rather their relative accuracy.

\begin{table}
\caption{\label{t:fiducial} Cosmological parameters for the
fiducial $\Lambda$CDM model around which the FMA is performed. We choose a
slightly open model to avoid numerical inaccuracies in the derivatives.}
\begin{ruledtabular}
\begin{tabular}{|l  c c|}
Baryons        & $\Omega_b$      & $0.046$  \\
Matter         & $\Omega_m$      & $0.270$  \\
Dark Energy    & $\OLa    $ & $0.720$  \\
Radiation      & $\Omega_{\textrm{rad}}$  & $7.95 \cdot 10^{-3}$ \\
Massless $\nu$ families & $N_\nu $         & 3.04 \\
Total density  & $\Omega_{\textrm{tot}}$  & $0.990$  \\
Hubble constant & $ h$           & $0.72$   \\
Optical depth  & $\treion$          & $0.166$  \\
Spectral index & $n_s$           & $0.99$ \\
Normalization  & $A_s$           & $2\cdot10^{-9}$
\end{tabular}
\end{ruledtabular}
\end{table}

\subsubsection{Forecasts and discussion}

Table~\ref{t:fmares} summarizes our forecasts for the future measurements
and compares them with the results obtained from WMAP actual data.

\begin{table}
\caption{\label{t:fmares} Fisher Matrix forecasts and
  comparison with present-day results, for different priors and
using different combinations of temperature and polarization CMB
spectra. Errors are in percent wrt the
values of the fiducial model, $Y_p=0.24$ and $\omega_b = 0.0238$ (
1-$\sigma$ c.l. all other marginalized).}
\begin{ruledtabular}
\begin{tabular}{|l | c c |  c c | c |}
\separator{Temperature, TE-cross, E-polarization}
      &  \multicolumn{2}{c|}{No priors} &
      \multicolumn{2}{c|}{Flatness} & Flatness and\\
      &  \multicolumn{2}{c|}{}      &   \multicolumn{2}{c|}{} & $Y_p = 0.24$ \\
      & $\frac{\Delta Y_p}{Y_p}$ & $\frac{\Delta \ob}{\ob}$
      & $\frac{\Delta Y_p}{Y_p}$ & $\frac{\Delta \ob}{\ob}$
      & $\frac{\Delta \ob}{\ob}$ \\
WMAP 4yrs  \footnotemark[1] \rule{0pt}{4ex}
      & $\sim50$ & $2.92$                 & $\sim40  $ & $2.86$                      & $2.86$     \\
Planck& $7.60$& $1.31$                 & $ 4.96$ & $1.26$                      & $0.70$    \\
CVL   & $2.59$& $0.34$                 & $ 1.52$ & $0.32$                      & $0.13$      \\\hline
\separator{Temperature + TE-cross}
WMAP 1st yr \footnotemark[2] \rule{0pt}{4ex}
      &  N/A  & N/A                    & $71.25$ & $5.04$                      & $5.04$     \\
WMAP 4yrs  \footnotemark[1] \rule{0pt}{4ex}
      & $\sim75$& $4.10$                  & $\sim60  $ & $3.94$                      & $3.94$     \\
Planck& $8.91$& $1.74$                 & $ 6.60$ & $1.63$                      & $0.74$    \\
CVL   & $5.18$& $0.55$                 & $ 2.84$ & $0.55$                      & $0.19$      \\
\end{tabular}
\end{ruledtabular}
\footnotetext[1]{~FMA forecast, 4 years mission including E-polarization.}
\footnotetext[2]{~Actual WMAP data and other CMB experiments, this work.}
\end{table}

We notice that when the WMAP full 4 years data will be available (including E-polarization),
the error on the
baryon density is expected to decrease by a factor of 2 to $2.86\%$, compared to today's
$5.04\%$ (assuming flatness).
Nevertheless, inclusion of $Y_p$ as a free parameter will still have
no effect on the determination of $\ob$ for WMAP, \ie $Y_p$ will remain
an essentially flat direction when marginalized over. While the determination of
the helium fraction will improve, the FMA cannot reliably assess quantitatively how much,
since for such large errors the likelihood distribution is not Gaussian and
the quadratic approximation breaks down. In the table we therefore give the
FMA estimation as an indication, with the caveat that the Fisher approximation is likely to
be inaccurate for the real errors on $Y_p$ from WMAP's 4 years data.

\begin{figure}[htb]
\begin{center}
\includegraphics[width=3.5in]{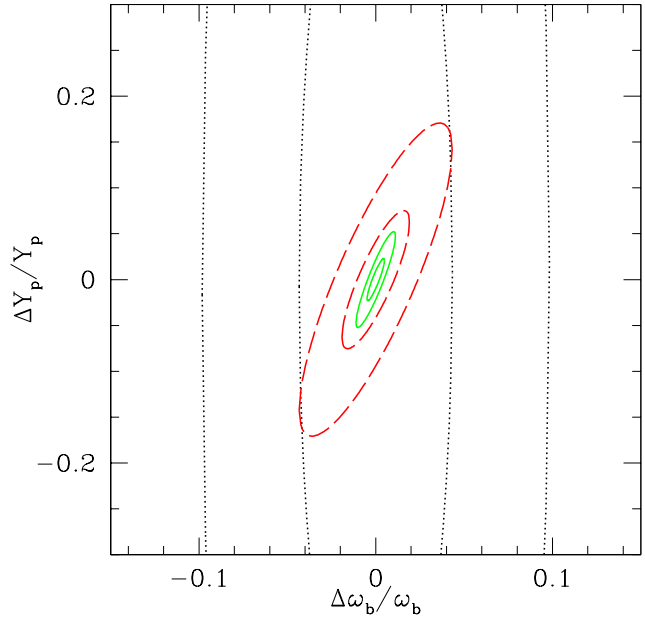}
\end{center}
\caption{FMA forecast for the expected errors from
WMAP 4 years mission (dotted-black), Planck (dashed-red) and a CVL
experiment (solid-green). The ellipses encompass 1-$\sigma$ and 3-$\sigma$
joint confidence regions for $\omega_b-Y_p$ (all other parameters marginalized).
The axis values give the error in wrt the fiducial model values.
This forecast is for the full CMB information (Temperature, TE-cross,
E-polarization) and
assumes flatness.}
\label{fig:fmaYpob}
\end{figure}

It is interesting that for Planck the effect of the helium
fraction can no longer be neglected. Inclusion of the helium
fraction increases the error on $\ob$ by roughly $80\%$, from
$0.70\%$ to $1.26\%$. The correlation between the two parameters
will have to be taken into account, as is evident from
Fig.~\ref{fig:fmaYpob}. The expected correlation coefficient is
$\text{corr}(Y_p, \omega_b) = 0.84\quad(0.91)$ for Planck (for
CVL). The expected $1-\sigma$ error on $Y_p$ is about $5\%$ for
Planck, or $\Delta Y_p \sim 0.01$. This is of the same order as
the spread in current astrophysical measurements. We conclude that
in Planck-accuracy data analysis it will be necessary to include
the uncertainty in the determination of the helium mass fraction,
at least in the form of a Gaussian prior over $Y_p$ of the type we
used in the CMB data analysis presented above.

Finally, measuring CMB temperature and polarization with cosmic
variance accuracy would allow to constrain $Y_p$ to within
$1.5\%$, or $\Delta Y_p \sim 0.0036$ (assuming flatness). Such an
ideal measurement would be able to discriminate between the
BBN-guided, deuterium based helium value and the current lowest
direct helium observations (cf.~Fig.~\ref{fig:compilation}).

Our forecasts for the uncertainty in the Helium mass fraction from
future observations are in excellent agreement with the findings
of Ref.~\cite{knox03}. There, the standard deviation on $Y_p$ for
Planck is estimated to be $\Delta Y_p = 0.012$. The authors of
Ref.~\cite{knox03} also consider an experiment (CMBPol) with
characteristics similar to our CVL, for which they forecast
$\Delta Y_p = 0.0039$, again in close agreement with our result.
An earlier work~\cite{eisenstein} found for Planck (temperature
and polarization) $\Delta Y_p = 0.013$, in satisfactory
concordance with our result. It should be noticed that the
forecast reported for MAP in Table 2 of Ref.~\cite{eisenstein},
namely $\Delta Y_p = 0.02$, is nothing but the Gaussian prior $Y_p
= 0.24 \pm 0.02$ which was assumed in their analysis.

The main source of improvement for the determination of $Y_p$ will
be the better sampling of the temperature damping tail provided by
Planck and the CVL. Polarization measurements have mainly the
effect of reducing the errors on other parameters. In fact, we
have checked that excluding from our FMA the $2 \leq \ell \leq 50$
region of the E-polarization and ET-correlation spectra changes
the forecast precision on $Y_p$ less than about 10-15\% for Planck
and less than a few percent for CVL. This supports the conclusion
that the low-$\ell$ reionization bump is not very useful in
measuring the helium abundance, because of the degeneracy with
$\zreion$.

\section{Conclusions}
\label{sec:conclusions}

We have analysed the ability of CMB observations to determine the
helium mass fraction, $Y_p$. We find that present data only allow a
marginal detection, $ 0.160 < Y_p < 0.501$ at 68\% c.l..  This
determination is completely independent from the usual astrophysical
observations and uses CMB data only.  We discuss degeneracies between
$Y_p$ and other cosmological parameters, most notably the baryon
abundance, the redshift and optical depth of reionization and the
spectral index. We conclude
that present-day CMB data accuracy does not require the inclusion of
$Y_p$ as a free parameter.
We find that Planck will
determine the helium mass fraction within $5\%$ (or $\Delta Y_p \sim
0.01$), which however will only allow a marginal discrimination
between different astrophysical measurements.
Nevertheless, we point out that the uncertainty of the helium fraction
will have to be taken into account
in order to correctly estimate the errors on the baryon density from Planck.
To determine if the emerging baryon tension (from BBN) is
related to underestimated systematic error-bars or whether it is an
indication of new physics, CMB observation will have to be pushed very close
to the cosmic variance limit in both temperature and polarization.

\section*{Acknowledgement}

It is a pleasure to thank Ruth Durrer, Lloyd Knox, Samuel Leach, Anthony Lewis,
Christophe Ringeval, Dominik Schwarz and Gary Steigman for useful comments and discussions. We thank the anonymous referee for many useful suggestions.
This work was performed
on the SUN Enterprise 10000 Supercomputer owned and operated by the
University of Geneva. R.T. is partially supported by the Swiss
National Science Fundation, the Schmidheiny Fundation
and the European Network CMBNET. S.H. thanks the Tomalla foundation
for support.


\end{document}